\documentclass[preprint,journal]{rmaa}

\usepackage{paralist}
\usepackage{{longtable}}
\usepackage{bm}
\usepackage{psfrag,color}
\usepackage[latin1]{inputenc}

\newcommand{\dechms}[4]{$#1^{\rm h}#2^{\rm m}#3\mbox{$^{\rm s}\mskip-7.6mu.\,$}#4$} %% \dechms{04}{14}{12}{9198} = RA en formato 04^h 14^m 12^s .9198
\newcommand{\decdms}[4]{$#1^{\circ}#2'#3\mbox{$''\mskip-7.6mu.\,$}#4$} %% \decdms{28}{12}{12}{199} = Dec en formato 28^o 12' 12'' .199
\newcommand{\msec}[2]{$#1\mbox{$''\mskip-7.6mu.\,$}#2$}
\newcommand{\mmsec}[2]{$#1\mbox{$^s\mskip-7.6mu.\,$}#2$}
\newcommand{\mmin}[2]{$#1\mbox{$'\mskip-7.6mu.\,$}#2$}
\title{Deep VLA observations of nearby star forming regions I: Barnard 59 and Lupus 1} 

\author{
  S. A. Dzib\altaffilmark{1},
  L. Loinard\altaffilmark{1,2},
  S.-N. X. Medina\altaffilmark{1},
  L. F. Rodr\'{\i}guez\altaffilmark{2},
  A. J. Mioduszewski\altaffilmark{3},
  and
  R. M. Torres\altaffilmark{4}
  }

\altaffiltext{1}{Max-Planck-Institut f\"ur Radioastronomie, Bonn, Germany.}
\altaffiltext{2}{Instituto de Radioastronom\'{\i}a y Astrof\'{\i}sica, Universidad Nacional Aut\'{o}noma de M\'{e}xico,
Apartado Postal 3-72, 58089 Morelia, Michoac\'{a}n, M\'exico.}
\altaffiltext{3}{National Radio Astronomy Observatory, Domenici Science Operations Center, 1003
Lopezville Road, Socorro, NM 87801, USA}
\altaffiltext{4}{Centro Universitario de Tonal\'a, Universidad de Guadalajara, Avenida Nuevo Perif\'erico
No. 555, Ejido San Jos\'e Tatepozco, C.P. 48525, Tonal\'a, Jalisco, M\'exico.}
\shortauthor{Dzib et al.}
\shorttitle{VLA observations to B59 and Lupus 1}

\fulladdresses{
\item S. Dzib: Max-Planck-Institut f\"ur Radioastronomie, Auf dem H\"ugel 69, 53121 Bonn, Germany (sdzib@mpifr-bonn.mpg.de).
\item L. Loinard: Instituto de Radioastronom\'{\i}a y Astrof\'{\i}sica, Universidad Nacional Aut\'{o}noma de M\'{e}xico,
Apartado Postal 3-72, 58089 Morelia, Michoac\'{a}n, M\'exico and Max-Planck-Institut f\"ur Radioastronomie, Auf dem H\"ugel
69, 53121 Bonn, Germany (l.loinard@crya.unam.mx)
\item S.-N. X. Medina: Max-Planck-Institut f\"ur Radioastronomie, Auf dem H\"ugel 69, 53121 Bonn, Germany (smedina@mpifr-bonn.mpg.de).
\item L. F. Rodr\'iguez: Instituto de Radioastronom\'{\i}a y Astrof\'{\i}sica, Universidad Nacional Aut\'{o}noma de M\'{e}xico,
Apartado Postal 3-72, 58089 Morelia, Michoac\'{a}n, M\'exico (l.rodriguez@crya.unam.mx).
\item A. J. Mioduszewski: National Radio Astronomy Observatory, Domenici Science Operations Center, 1003
Lopezville Road, Socorro, NM 87801, USA (amiodusz@nrao.edu).
\item R. M. Torres: Centro Universitario de Tonal\'a, Universidad de Guadalajara, Avenida Nuevo Perif\'erico
No. 555, Ejido San Jos\'e Tatepozco, C.P. 48525, Tonal\'a, Jalisco, M\'exico (rosamarthatorres@gmail.com).
}

\listofauthors{Author1, Author2, \& Author3}
\indexauthor{Author1, W. J.}
\indexauthor{Collaborator, A.}
\indexauthor{Author, L.}

\resumen{Barnard 59 y Lupus 1 son dos regiones de formaci\'on estelar cercanas, visibles desde el hemisferio sur. En este manuscrito,  presentamos observaciones profundas ($\sigma$ $\lesssim$ 15 $\mu$Jy) en radio frecuencias ($\nu$ = 6 GHz; $\lambda$ = 5 cm) hacia estas dos regiones, y reportamos la detecci\'on de un total de 114 radio-fuentes. Trece de estas fuentes est\'an asociadas con objetos estelares j\'ovenes previamente conocidos, nueve en Barnard 59 y cuatro en Lupus 1. Las caracter\'{\i}sticas de la emisi\'on radio (el \'{\i}ndice espectral y, en algunos casos, la polarizaci\'on) sugieren que la emisi\'on radio tiene un origen t\'ermico en la mayor\'{\i}a de los objectos estelares j\'ovenes. Solamente para dos fuentes (Sz~65 y Sz~67), encontramos indicios de un posible origen no-t\'ermico; m\'as observaciones ser\'an necesarias para constre\~nir la naturaleza de la emisi\'on radio en estos dos casos. Las dem\'as fuentes detectadas en radio no tienen contrapartes en otras longitudes de onda y el n\'umero de fuentes detectadas por unidad de \'angulo s\'olido coincide con el esperado para fuentes de fondo. Esto sugiere que todas las fuentes radio que no est\'an asociadas con objetos estelares j\'ovenes conocidos son fuentes extragal\'acticas de fondo.}

\abstract{Barnard 59 and Lupus 1 are two nearby star-forming regions visible from the southern hemisphere. In this manuscript, we present deep ($\sigma$ $\lesssim$ 15 $ \mu$Jy) radio observations ($\nu$ = 6 GHz; $\lambda$ = 5 cm) of these regions, and report the detection of a total of 114 sources. Thirteen  of these sources are associated with known young stellar objects, nine in Barnard 59 and four in Lupus 1. The properties of the radio emission (spectral index and, in some cases, polarization) suggest a thermal origin for most young stellar objects. Only for two sources (Sz~65 and Sz~67) are there indications for a possible non-thermal origin; more observations will be needed to ascertain the exact nature of the radio emission in these sources. The remaining radio detections do not have counterparts at other wavelengths, and the number of sources detected per unit solid angle is in agreement with extragalactic number counts. This suggests that all radio sources not associated with known young stellar objects are background extragalactic sources.}

\addkeyword{radio continuum: stars}
\addkeyword{radiation mechanisms: non-thermal}
\addkeyword{radiation mechanisms: thermal}
\addkeyword{techniques: interferometric}

\begin{document}

\maketitle

\section{INTRODUCTION}
\label{sec:intro}
Low-mass young stellar objects (YSOs) are newly born stars that are not yet burning hydrogen in their cores. Most of them are embedded in their parental molecular clouds and are, therefore, obscured at optical wavelengths. At radio wavelengths ($\nu$ $\lesssim$ 30 GHz) the cloud is practically transparent and YSOs with radio emission can be studied (e.g.\ Dulk 1985; G\"udel 2002). The upgrades recently implemented on several large radio interferometers (e.g.\ the Karl G. Jansky Very Large Array; VLA) have revived this field of reseach by enabling large areas to be mapped in reasonable amounts of time. For instance, the Gould's Belt Very Large Array Survey has revealed a large population of radio-bright YSOs in the five most often studied nearby star-forming regions: Ophiuchus (Dzib et al.\ 2013a), Orion (Kounkel
et al.\ 2014), Perseus (Pech et al.\ 2015), Serpens (Ortiz-Le\'on et al.\ 2015), and Taurus (Dzib et al.\ 2015). 

The present paper is the first in a new series reporting on radio observations of additional nearby regions of star formation (at $d$ $\lesssim$ 1 kpc). The goal of these observations is to detect and characterize the radio emission from {known} YSOs. Thus, these data will provide a fairly complete inventory of radio-emiting YSOs, and will constrain the radio emission mechanisms. Our results will be compared with those previously found in other regions. The present manuscript will focus on two nearby star-forming regions visible from the southern hemisphere:  Barnard~59 (B59) and Lupus~1.

\subsection{Barnard 59 and Lupus 1}

\begin{figure*}[t!]
\begin{center}
\begin{tabular}{cc}
\includegraphics[width=0.49\textwidth,trim= 120 10 192 0, clip]{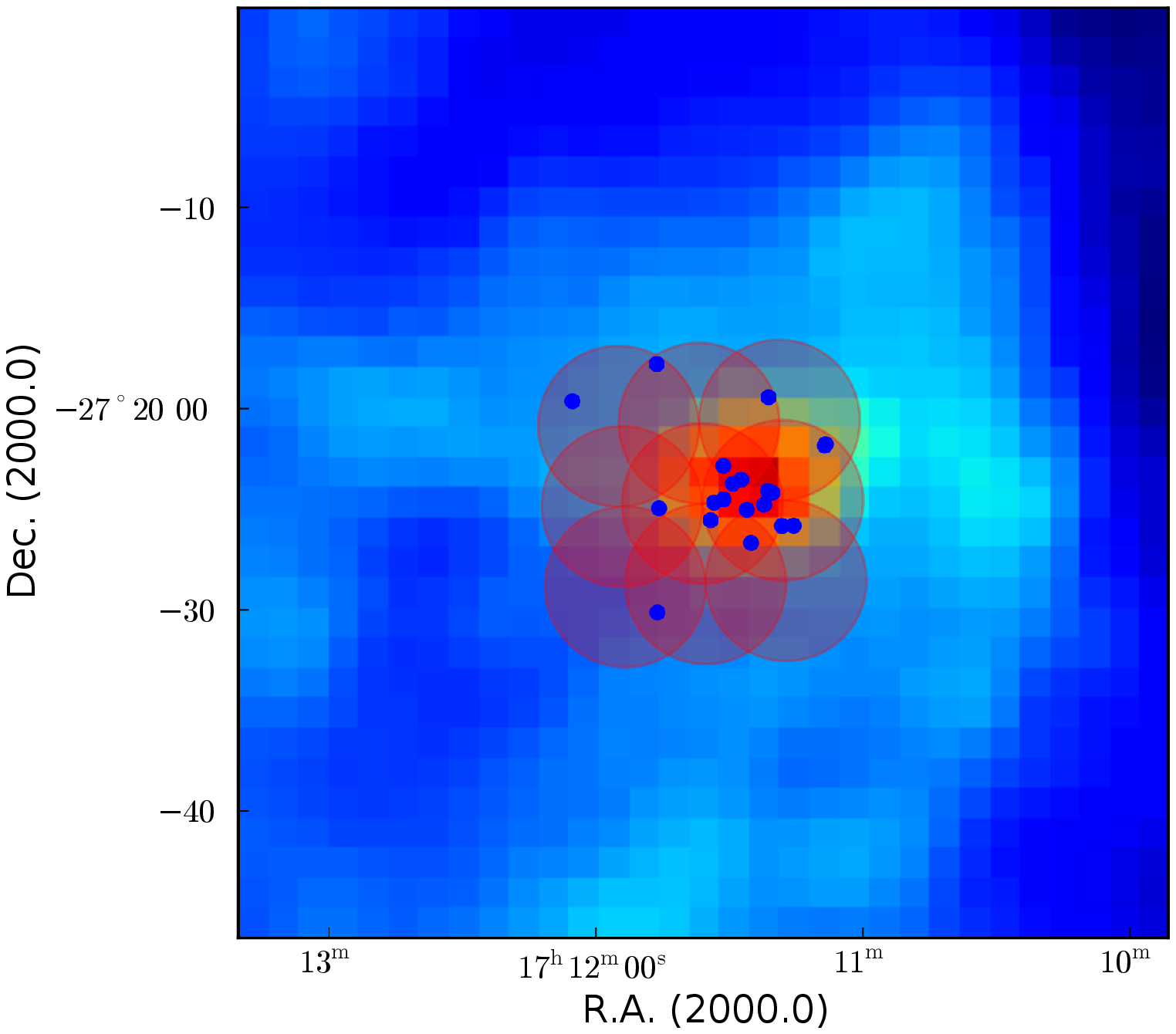} &
\includegraphics[width=0.49\textwidth,trim= 122 10 190 0, clip]{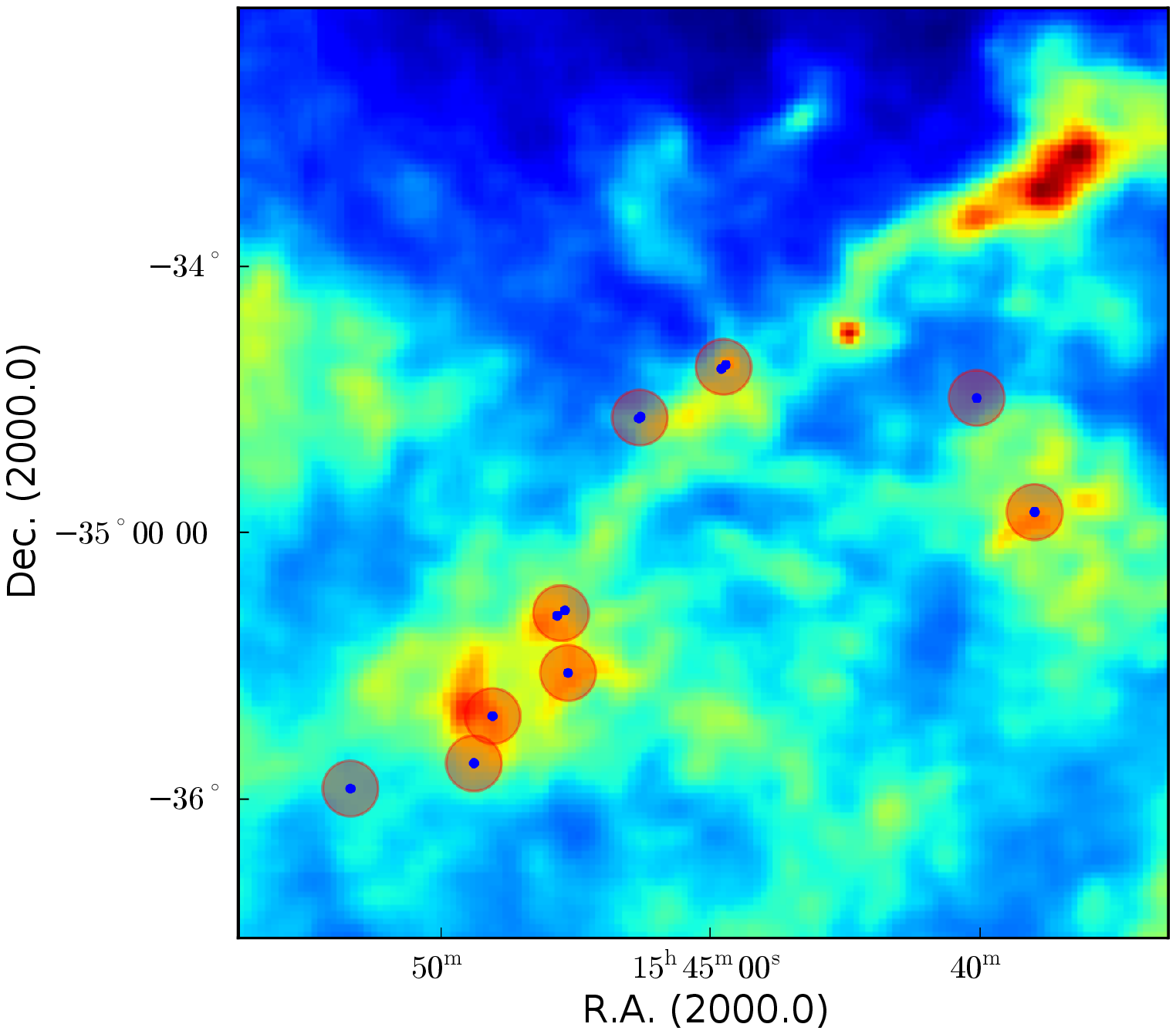}
\end{tabular}
\end{center}
\caption{Mapped areas in B59 (left) and Lupus 1 (right). The red circles indicate the observed VLA fields. The diameter of each circle is 8$'$ in B59 and \mmin{12}{5} in Lupus 1 (see text). Blue dots are known YSOs. Background: IRAS images at 100 $\mu$m of the B59 and Lupus 1 region (Neugebauer et al.\ 1984). }
\label{fig:b59}
\end{figure*}

The Pipe nebula is a large molecular complex composed of several dark clouds (Barnard 1919; Lynds 1962). Among those, B59 stand out as the only one with significant star formation activity (Alves, Lombardi \& Lada 2008). There are 20 catalogued young stars in B59 in an area of 
$15'\times15'$, of which 15 are T Tauri star candidates (Brooke et al.\ 2007). Recently, Dzib et al.\ (2013b) reported on radio observations that covered the central region of B59 and detected five radio sources associated with five of the YSOs reported from the Spitzer observations of Brooke et al.\ (2007). These are thought to be thermal radio sources because of their SED classification and the $L_{\rm bol}-L_{\rm radio}$ ratio, which agrees well with that expected from the correlation of free-free radio continuum and the stellar luminosity (e.g., AMI Consortium et al.\ 2011). The observations presented here expand upon those reported by Dzib et al.\ (2013b) and cover the 20 YSOs reported by Brooke et al.\ (2007).

In his recent review, Comer\'on (2008) recognized Lupus as one of the main low-mass star-forming-region complexes within 200 pc of the Sun. The complex is divided into nine individual clouds, labeled as Lupus 1 to 9 (i.e., Hara et al.\ 1999), and they extend down to declinations as low as $-44^\circ$. The present study is restricted to the higher declination cloud (Lupus 1) at $-35^\circ$. Lupus 1 was the first cloud in the Lupus region where T Tauri stars were discovered (Henize 1954). Nowadays, up to a total of 13 classical T Tauri stars are known to be present in Lupus 1 (Comer\'on 2008), and some additional candidates have been suggested by Mer\'{\i}n et al.\ (2008) using the Spitzer Space Telescope. These YSOs are sparsely distributed suggesting that an isolated mode of star formation is at work. The observations presented here will cover the 13 known T Tauri stars in Lupus 1.

\section{OBSERVATIONS AND SOURCE SELECTION}
\label{sec:obs}

The observations were collected on 2014 March 6 (Lupus 1) and 20 (B59) with the VLA of the National Radio Astronomy Observatory (NRAO\footnote{The National Radio Astronomy Observatory is operated by Associated Universities Inc. under cooperative agreement with the National Science Foundation.}) in its most extended (A) configuration. The C band receiver was used and four frequency sub-bands, each 1 GHz wide and centered at 4.5, 5.5, 6.5, and 7.5 GHz, respectively, were recorded simultaneously with 3-bit sampling. For each region, nine target fields were observed. The fields in B59 were distributed around the core in a semi-mosaic configuration (Figure \ref{fig:b59}; left panel), while the fields in Lupus 1 were distributed to observe known YSOs (Figure \ref{fig:b59}; right panel). The phase calibrators were J1700-2610 and J1607-3331 for B59, and Lupus 1, respectively.

The total observing time per epoch was one hour. At the beginning of the observations, the first 10 minutes were spent on the flux calibrator 1331+305 = 3C286; part of this scan was spent slewing the antennas and setting up the correlator. Subsequently, the phase calibrator was observed by five minutes (again part of this time was spent on slewing the antennas). Finally, a cycle with three succesive target scan (each 4.5 minutes long) and one phase calibrator scan (1 minute) was repeated until all the targets were observed.

The data were edited and calibrated using the Common Astronomy Software Applications (CASA v4.2.2) package, using the VLA calibration pipeline v1.3.1. The bootstrapped flux density at $\nu_{\rm central}$ = 6.0 GHz for the phase calibrator J1700-2610 was 1.96\,$\pm$\,0.02 Jy, with a spectral slope $\alpha=0.10$. For J1607-3331, the bootstrapped flux density (also at 6 GHz) was 0.168\,$\pm$\,0.002 Jy, with $\alpha=-0.41$.

To obtain the highest sensitivity for the target fields, the four sub-bands were jointly imaged using a multi-frequency deconvolution (e.g., Rau \& Cornwell 2011). Additionally, in B59 we combined the fields in mosaic mode (imagermode='mosaic' in CASA) to gain further sensitivity. The pixel size was set to \msec{0}{06}, and a weighting scheme intermediate between natural and uniform was used (robust=0.0 in CASA).  { Also, the minimum level of the primary beam was set and corrected at a response of 20\% (PBCOR=True and MINPB=0.2 in CASA). The resulting mosaic in B59 covered an area of 316 arcmin$^2$. In the case of Lupus 1, each field covers a circular area with a diameter of \mmin{12}{5}.} The parameters of the maps are given in Table~\ref{tab:obs}.

A visual inspection of the images was first performed at the position of known YSOs to look for detections. { We consider a radio source to be the counterpart of a YSO when it has a flux density larger than four times the noise level in the images, and coincides with the position of  a previously reported YSO to within 1$''$. This corresponds to our angular resolution. Furthermore, an automated search for additional radio sources present in the mapped areas was performed using a specialized software, as we now describe.

An important element in the process of source extraction is to have the correct noise map. The rms estimation algorithm of the sExtractor package (Bertin \& Arnouts 1996) was used to create a suitable noise map. This algorithm defines the rms value for each pixel in an
image by determining the distribution of pixel values within a surrounding local mesh until all values are around $\pm3\times\sigma_{\rm rms}$. This method ensures that most real emission is removed from the noise image and the determined noise map contains the correct noise level. Once the rms map has been constructed, the source extraction was done using the BLOBCAT package (Hales et al.\ 2012). This software is written in the scripting language Python and utilizes the flood fill algorithm to detect and catalog blob sources in two dimensional astronomical images. It separates individual blobs from a dimensionless signal-to-noise (SNR) map. The SNR map is the result of a ratio between the input image and the given rms map. The extraction is performed, in the SNR map, searching all pixels and their neighbors, with a $5\times\sigma_{\rm rms}$ threshold. In this case, only the sources with integrated fluxes and peak fluxes above 5$\times\sigma_{\rm noise}$ were considered as real, { this allows us to weed out noise peaks from real sources}.}

{ It is worth explaining briefly why a threshold of 4$\sigma$ was chosen for radio sources associated with known YSOs, whereas 5$\sigma$ was required for sources with no known counterparts. As explained in detail in Pech et al.\ (2016), in an image where the noise level is $\sigma$ and where no sources at all are present, there is a probability of 2.87$\times$10$^{-7}$ that any given independent pixel will have emission at 5$\sigma$ or more. We emphasize that the presence of such a 5$\sigma$ peak is entirely due to random noise fluctuations, and not to a real source. As mentioned earlier, the area of our B59 mosaic is 316 arcmin$^2$, and our resolution element (\msec{0}{78} $\times$ \msec{0}{26}) has an area of 5.63$\times$10$^{-5}$ arcmin$^2$. Thus, there are 5.6$\times$10$^6$ independent resolution elements in our mosaic, and we expect at most (5.6$\times$10$^6$) $\times$ (2.87$\times$10$^{-7}$) = 1.6 noise peaks at 5$\sigma$ or more in our image. Thus, most of the sources detected at more than 5$\sigma$ must be real sources rather than noise peaks. We note that if we had chosen 4$\sigma$ instead, the expected number of false-positive detections would have increased to the completely unacceptable figure of 178.  In the case of our search around known YSOs, the search area is limited to 1 arcsec$^2$ (see above), which corresponds to barely 5 resolution elements. The probability of encountering a 4$\sigma$ peak in such a small area is extremely small (less than 0.02\%), so any source detected at more than 4$\sigma$ is very likely real.}

\begin{table}[!th]
\footnotesize
  \begin{center}
  \caption{ Observations and final parameters of maps.}
    \begin{tabular}{ccccccrc}\hline\hline
Region  &   Date  & Synthesized beam  & rms noise \\
   &(dd.mm.yy)   &($\theta_{\rm maj}\times\theta_{\rm min};$ P.A.) & ($\mu$Jy bm$^{-1}$)\\
   \noalign{\smallskip}
\hline\hline\noalign{\smallskip}
%%%%%
 B59 & 20.03.14 &\msec{0}{78}$\times$\msec{0}{26};\, 28.8$^\circ$  & 9\\
 Lupus 1  & 06.03.14 &\msec{0}{87}$\times$\msec{0}{24};\, 4.9$^\circ$ & 15\\
\hline\hline\noalign{\smallskip}
%\tabnotetext{a}{For individual fields.}
 \label{tab:obs}
%%%%%
\end{tabular}
\end{center}
\end{table}

\section{Results}
\label{sec:Res}

\subsection{Radio sources in B59}

A total of 56 sources were detected in the B59 region, nine of them associated with known YSOs (their derived parameters are given in Table~\ref{tabB59}). Five of these sources were previosuly detected at radio wavelengths by Dzib et al. (2013b) while the other four are reported here for the first time. The remaining sources are most likely associated with extragalactic objects and are discussed below. The deconvolved size of most of the detected YSOs are consistent with a point source structure. The only exceptions are $[$BHB2007$]$~11 with a deconvolved size of { (\msec{0}{43}$\pm$\msec{0}{07}$\times$\msec{0}{37}$\pm$\msec{0}{08}; PA=170$^\circ$$\pm$46$^\circ$), and $[$BHB2007$]$~10 with  a deconvolved size of (\msec{0}{50}$\pm$\msec{0}{25}$\times$\msec{0}{24}$\pm$\msec{0}{06}; PA=35$^\circ$$\pm$09$^\circ$).}

\begin{figure*}[ht!]
\begin{center}
\begin{tabular}{ccc}
\includegraphics[width=0.32\textwidth,trim= 10 0 30 0, clip]{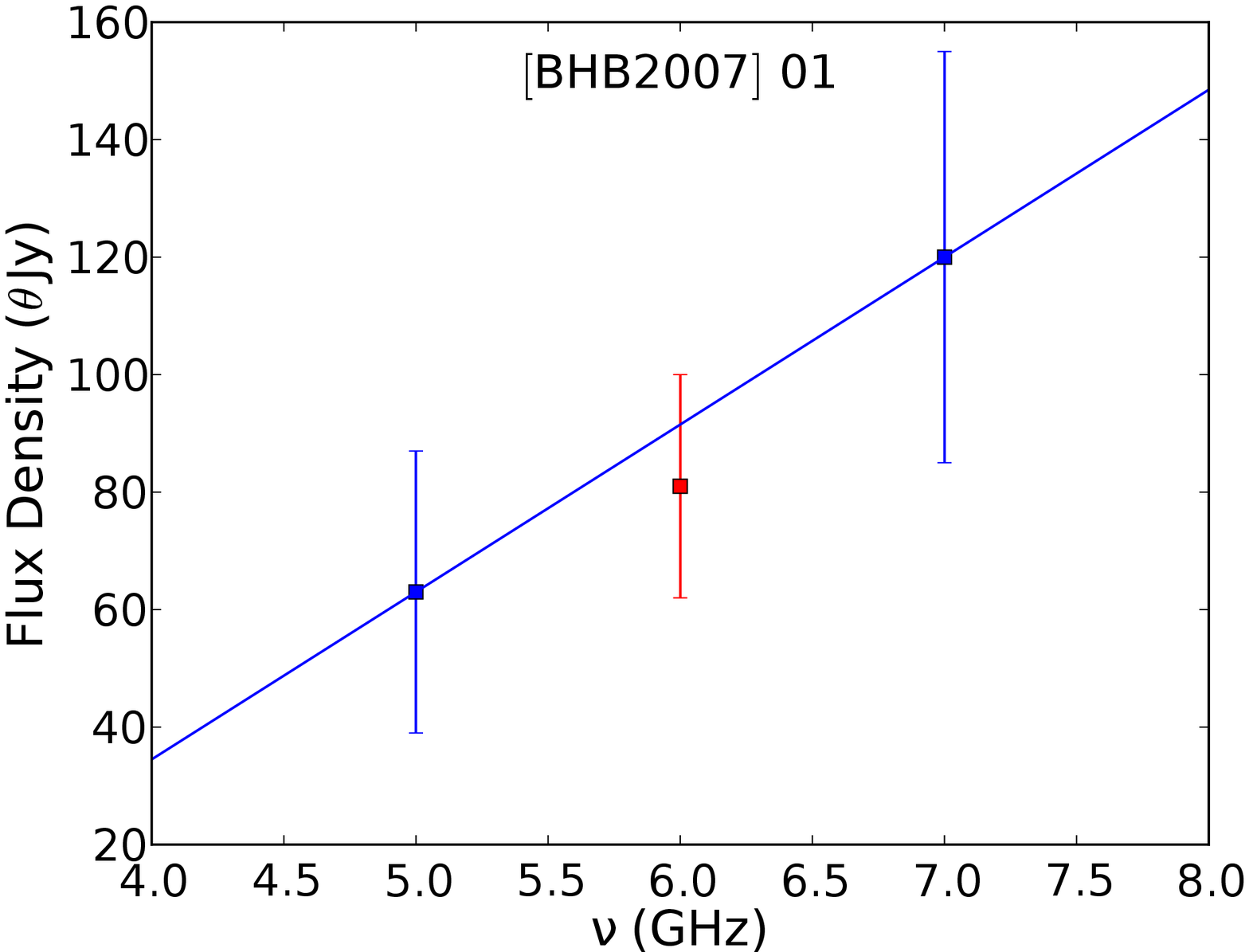} &
\includegraphics[width=0.32\textwidth,trim= 10 0 30 0, clip]{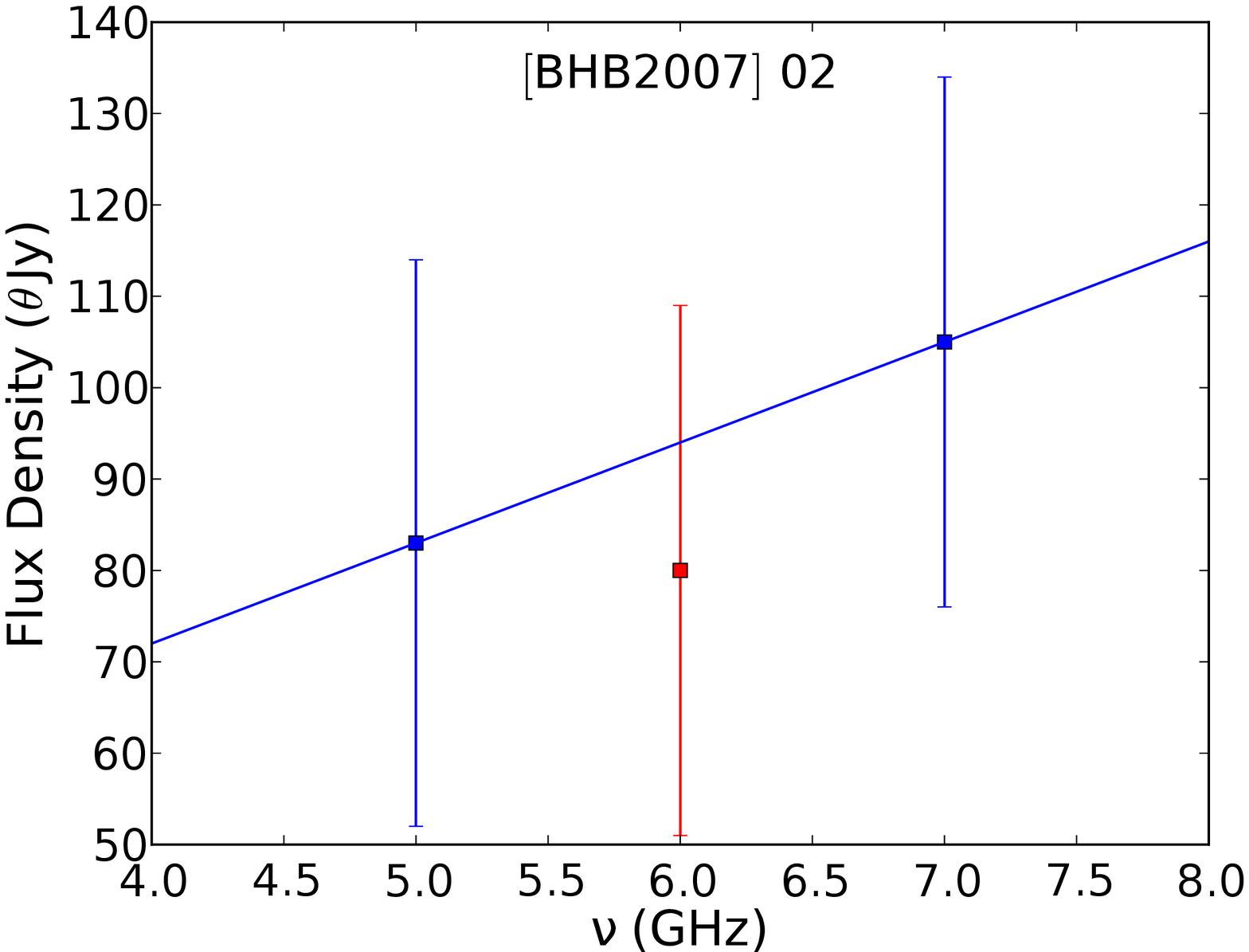} &
\includegraphics[width=0.32\textwidth,trim= 10 0 30 0, clip]{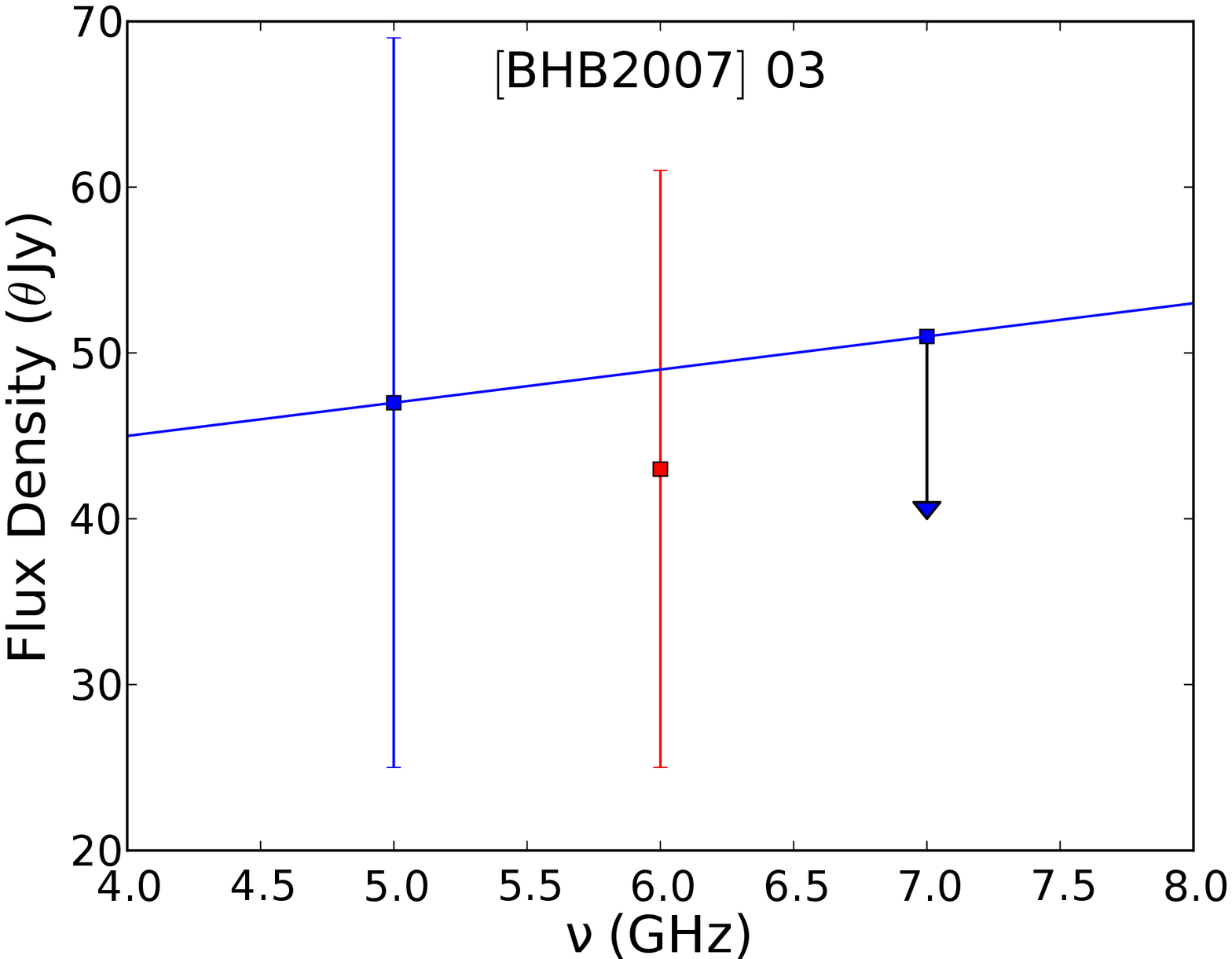} \\
\includegraphics[width=0.32\textwidth,trim= 10 0 30 0, clip]{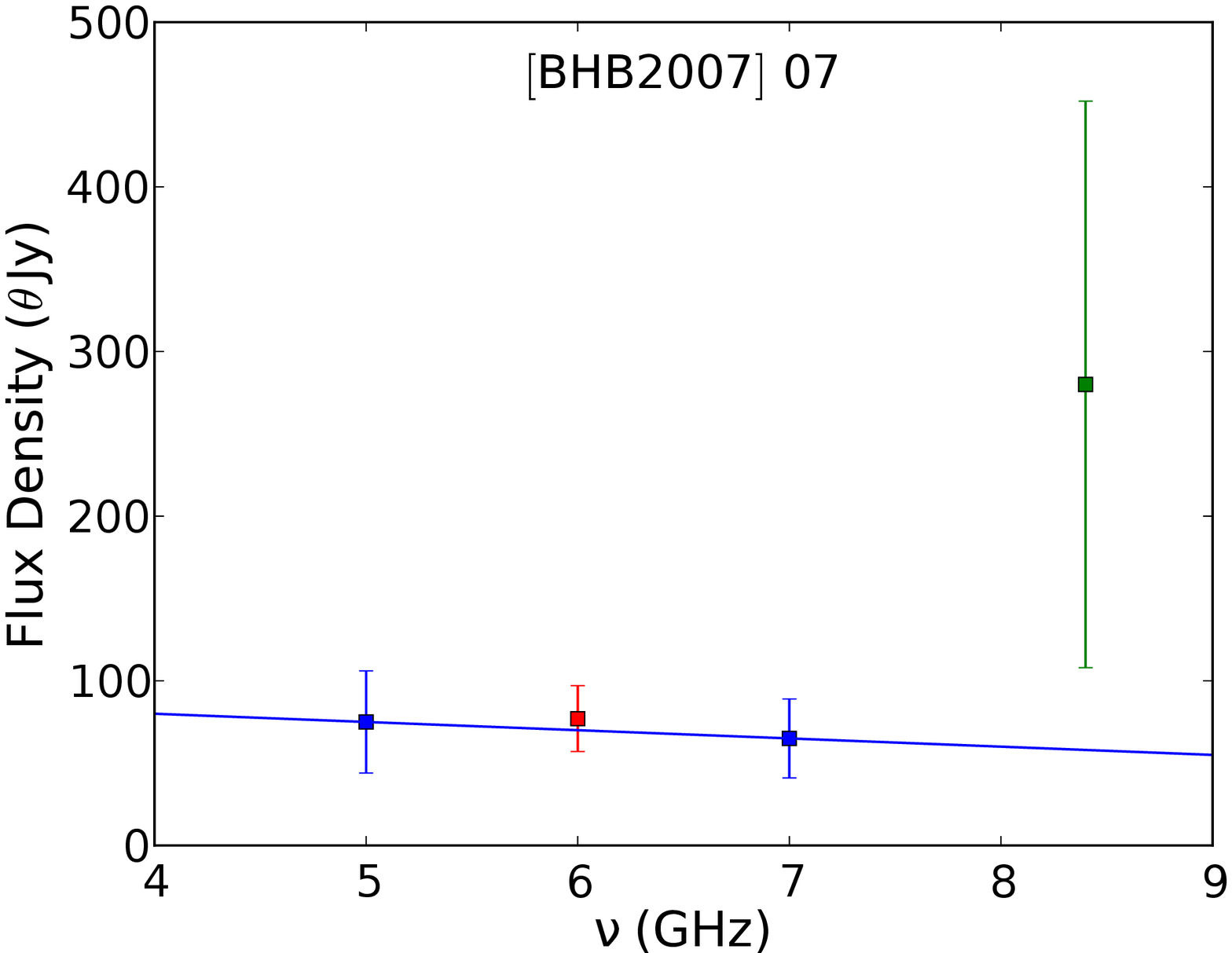} &
\includegraphics[width=0.32\textwidth,trim= 10 0 30 0, clip]{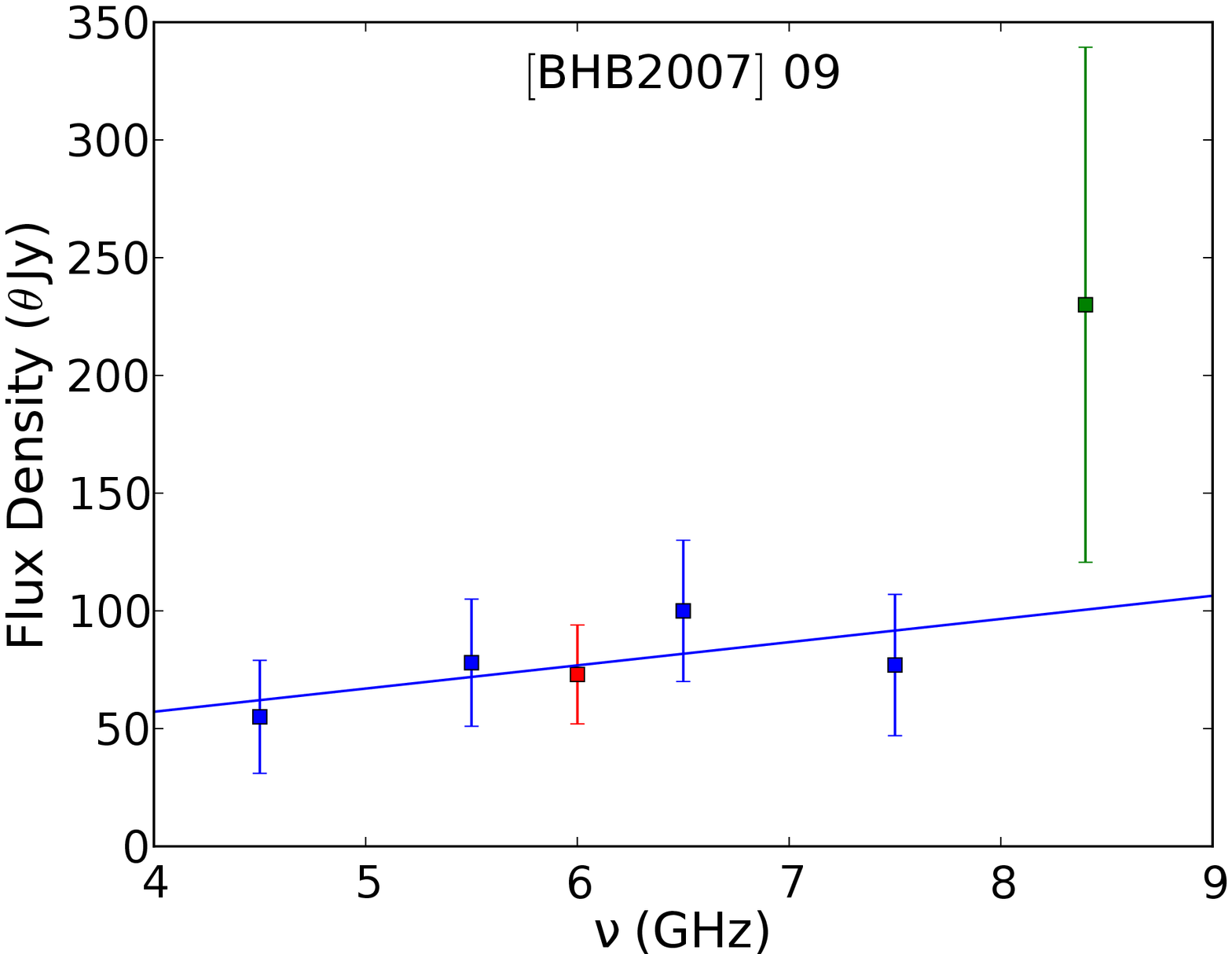} &
\includegraphics[width=0.32\textwidth,trim= 10 0 30 0, clip]{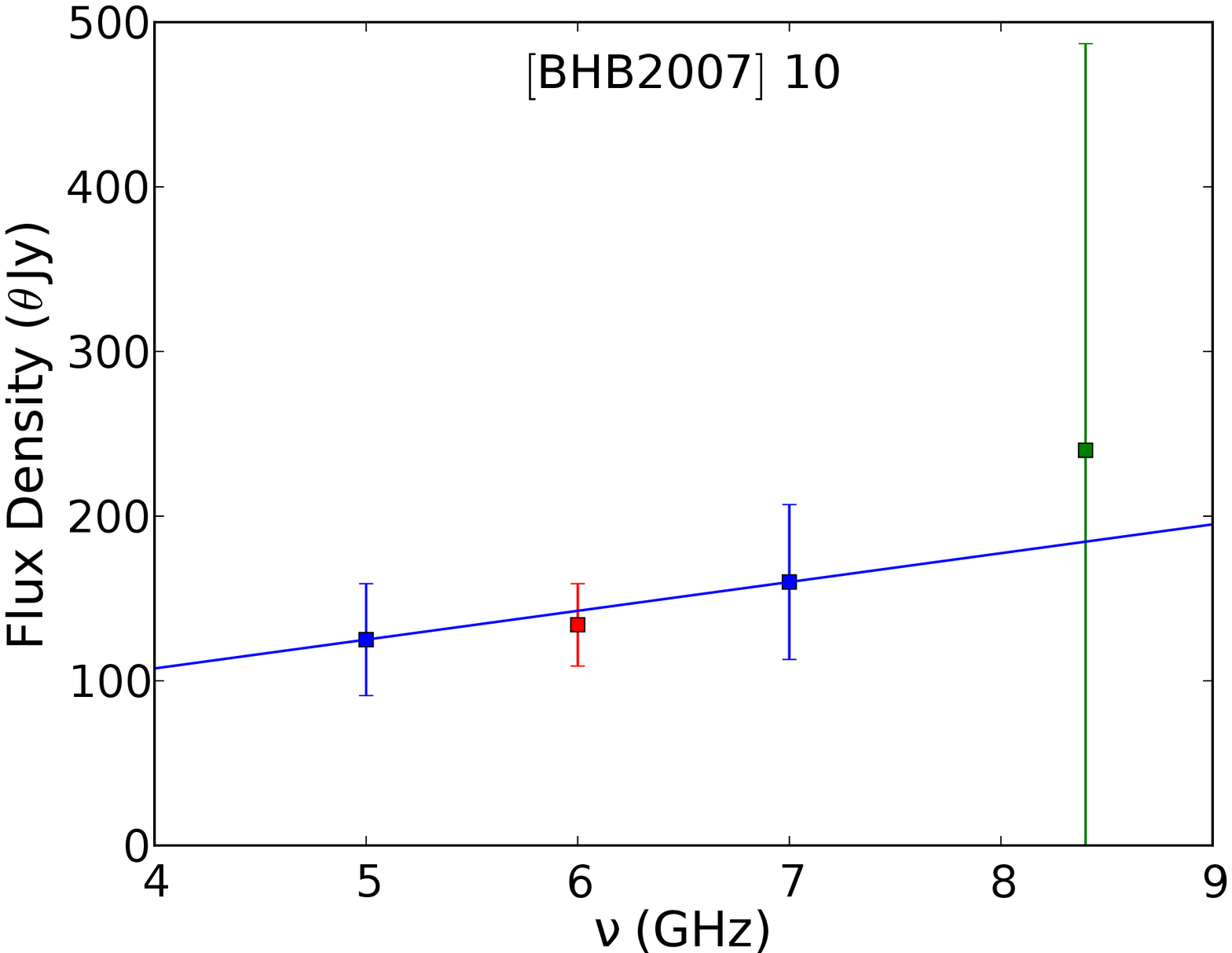} \\
\includegraphics[width=0.32\textwidth,trim= 10 0 30 0, clip]{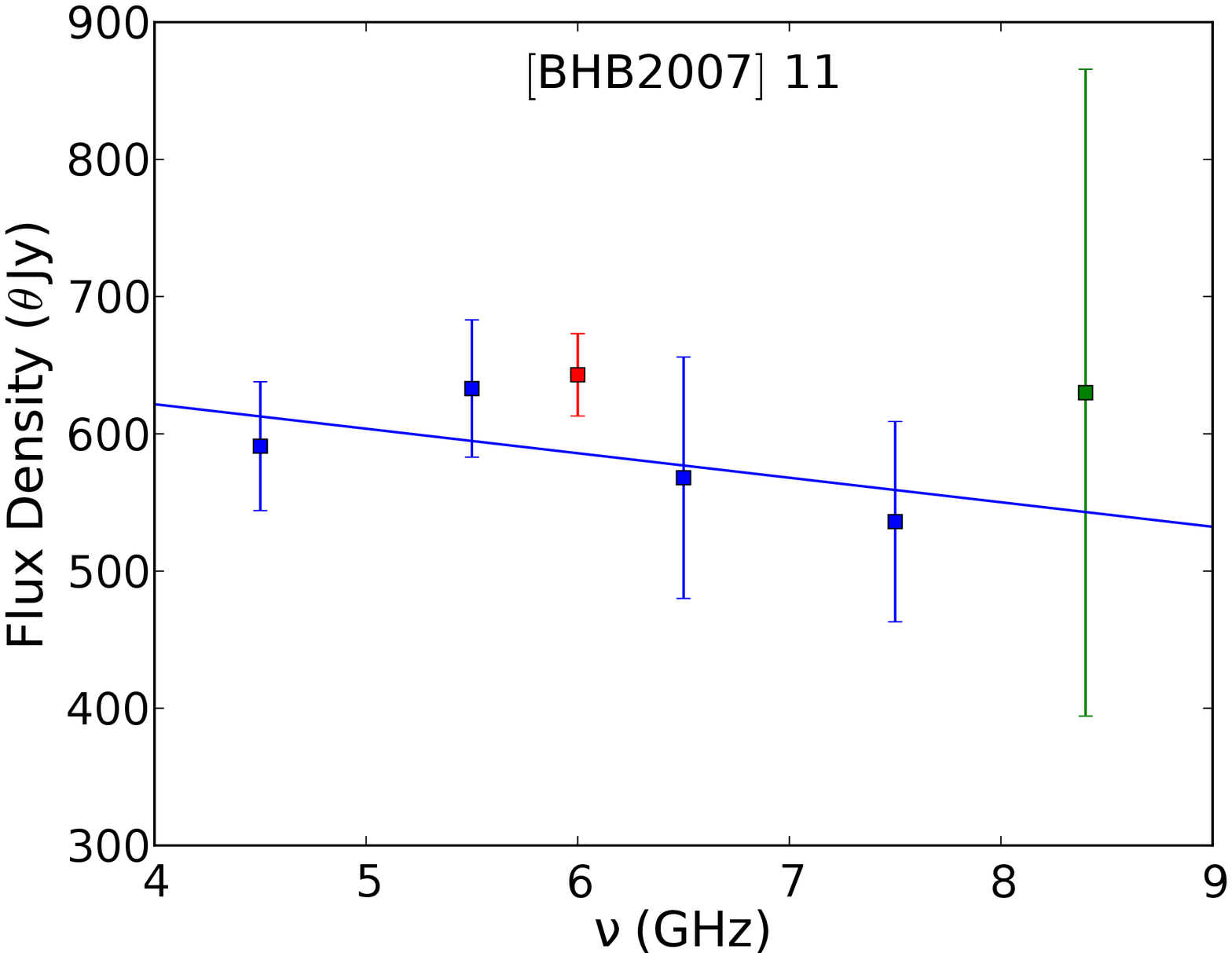} &
\includegraphics[width=0.32\textwidth,trim= 10 0 30 0, clip]{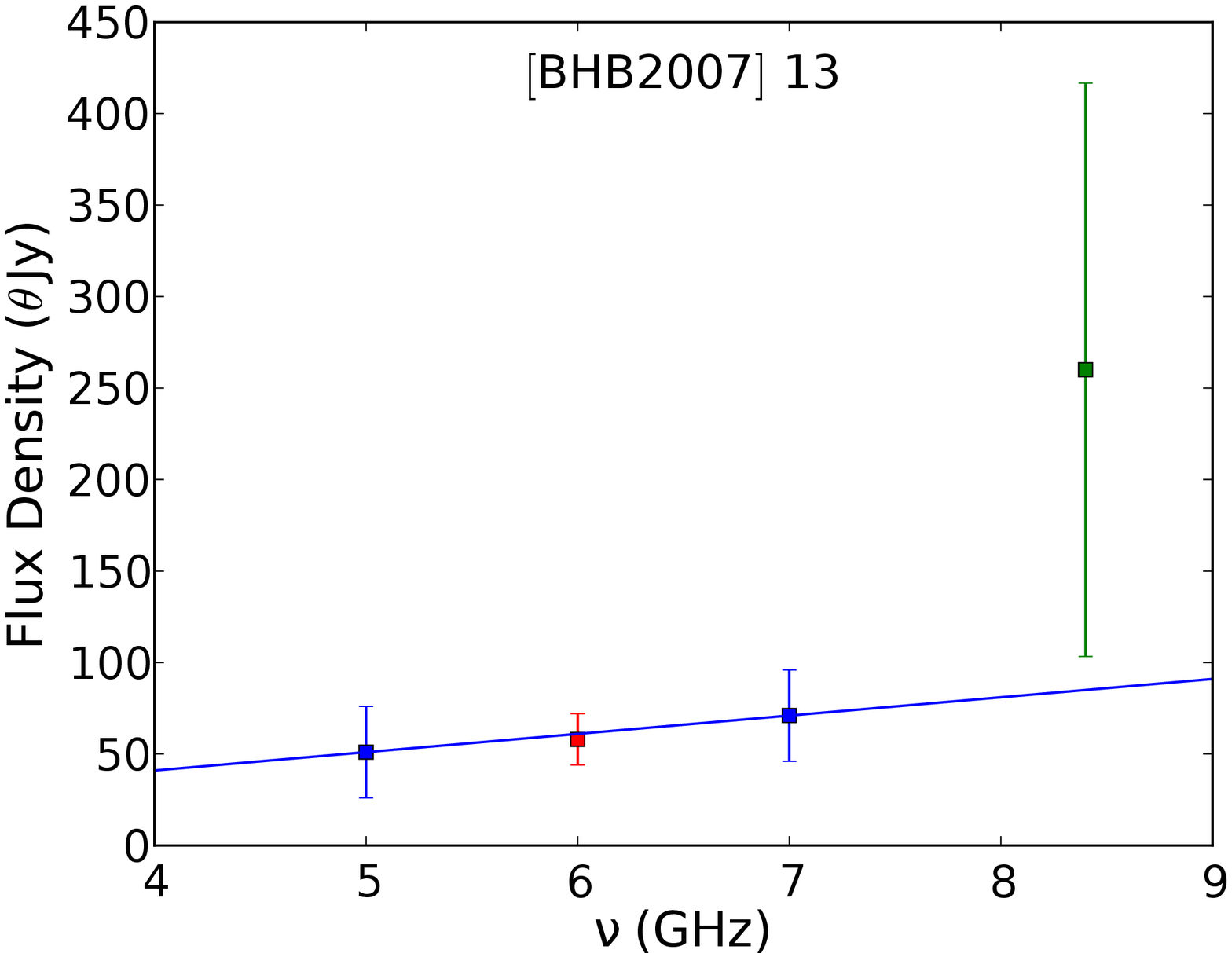} &
\includegraphics[width=0.32\textwidth,trim= 10 0 30 0, clip]{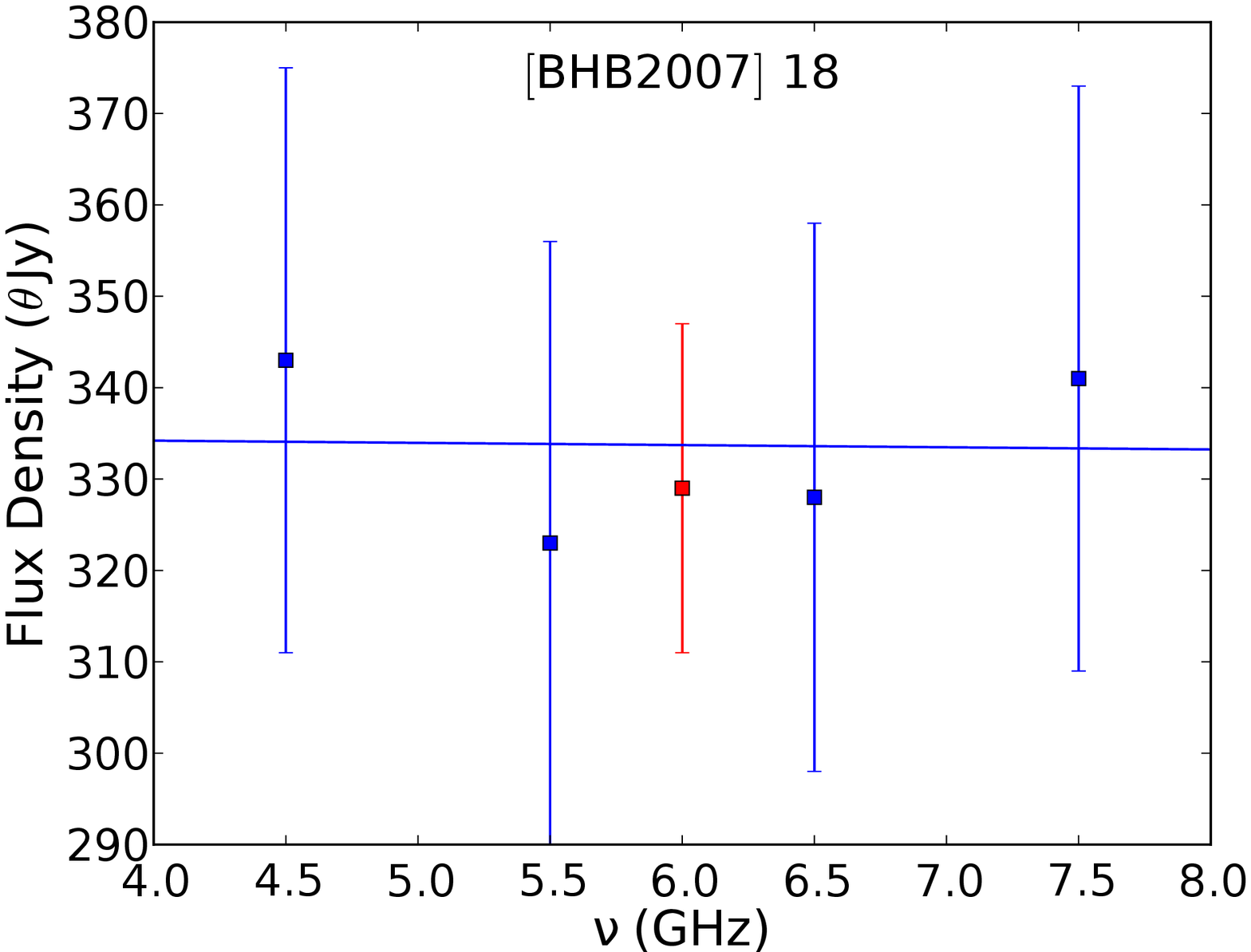} 
\end{tabular}
\end{center}
\caption{Spectral index fitting for the sources detected in B59, using the images obtained in independent sub-bands (blue points). The red points at 6~GHz are the flux densities of four combined sub-bands. Green points are the fluxes measured by Dzib et al.\ (2013b) at 8.4~GHz, and their error bars are the result of adding in quadrature the flux errors as reported by Dzib et al.\ (2013b) and the uncertainties for the spectral index fitting.}
\label{fig:b59_a}
\end{figure*}

As a first attempt to determine the spectral index, $\alpha$ (defined as $S_\nu\propto\nu^{\alpha}$) of the detected YSOs, the data sets were split and imaged in several sub-bands. For the brightest sources, the data were split in four separate sub-bands, of 1~GHz each. For the weaker sources, only two sub-bands, of 2~GHz each, were imaged. The spectral index was determined from least-square fitting to the flux densities measured in the detected sub-bands. The resulting spectral indices are given in Table~\ref{tabB59} and the 
fits are shown in Figure~\ref{fig:b59_a}. A circular polarization search was not possible since all detected YSOs in B59 are far from the observed phase center and the beam squint effect\footnote{This effect is produced by the separation of the R and L beams on the sky, and heavily affects sources far from the field center.} would stronly affect the results.

\begin{table*}[!th]\centering
  \setlength{\tabnotewidth}{2.0\columnwidth}
  \tablecols{6}
  % Stretch the space between table columns 
  %\setlength{\tabcolsep}{2.8\tabcolsep}
  \caption{Radio properties of YSOs in B59} \label{tabB59}
 \begin{tabular}{lccccc}
    \toprule
YSO & SED  & \multicolumn{2}{c}{ VLA Position} & { $S_\nu\pm\sigma_{S_\nu}$\tabnotemark{a} }& ${\alpha}$\\
Name& Class&  RA    &   Dec.    &   ($\mu$Jy) & \\
       \midrule
$[$BHB2007$]$ 1 & Flat & \dechms{17}{11}{03}{{937}}$\pm$\mmsec{0}{003} & \decdms{-27}{22}{55}{26}$\pm$\msec{0}{06} & 81$\pm$19&${\quad1.9\pm1.4}$\\%
$[$BHB2007$]$ 2 & II & \dechms{17}{11}{04}{{137}}$\pm$\mmsec{0}{004}& \decdms{-27}{22}{59}{39}$\pm$\msec{0}{09} & 80$\pm$29&${\quad0.7\pm1.4}$\\%
$[$BHB2007$]$ 3 & II & \dechms{17}{11}{11}{{826}}$\pm$\mmsec{0}{003}& \decdms{-27}{26}{55}{10}$\pm$\msec{0}{10} & 43$\pm$18&${<0.2\pm1.7}$\\%
$[$BHB2007$]$ 4 & II &... & ... & $<$36 &...\\
$[$BHB2007$]$ 5 & II &... & ... & $<$36 &...\\
$[$BHB2007$]$ 6 & II &... & ... & $<$36 &...\\
$[$BHB2007$]$ 7 & Flat & \dechms{17}{11}{17}{{286}}$\pm$\mmsec{0}{003} &\decdms{-27}{25}{08}{36}$\pm$\msec{0}{06} & 77$\pm$20&${-0.4\pm1.6}$\\%
$[$BHB2007$]$ 8 & II &... & ... & $<$36 &...\\
$[$BHB2007$]$ 9 & Flat & \dechms{17}{11}{21}{{553}}$\pm$\mmsec{0}{002} &\decdms{-27}{27}{42}{17}$\pm$\msec{0}{04}& 73$\pm$21&${\quad0.8\pm1.0}$\\%
$[$BHB2007$]$ 10 &I &  \dechms{17}{11}{22}{{161}}$\pm$\mmsec{0}{003} &\decdms{-27}{26}{ 02}{02}$\pm$\msec{0}{06} & 135$\pm$24&${\quad0.7\pm1.2}$\\% 503 +-246mas  242 +-56 mas   35 +-9d
$[$BHB2007$]$ 11 &0/I &  \dechms{17}{11}{23}{{118}}$\pm$\mmsec{0}{001}& \decdms{-27}{24}{32}{56}$\pm$\msec{0}{02} & 643$\pm$30&${-0.1\pm0.3}$\\%298+-15  430+-110 mas 374+-138 mas 176+-60d
$[$BHB2007$]$ 12 &II & ... & ... & $<$36 &...\\
$[$BHB2007$]$ 13 &II &  \dechms{17}{11}{27}{{017}}$\pm$\mmsec{0}{002} &\decdms{-27}{23}{48}{68}$\pm$\msec{0}{06} & 58$\pm$14&${\quad1.0\pm1.8}$\\%
$[$BHB2007$]$ 14 &II & ... & ... & $<$36 &...\\
$[$BHB2007$]$ 15 &II & ... & ... & $<$40 &...\\
$[$BHB2007$]$ 16 &II & ... & ... & $<$40 &...\\
$[$BHB2007$]$ 17 &II & ... & ... & $<$36 &...\\
$[$BHB2007$]$ 18 &II &  \dechms{17}{11}{41}{{8402}}$\pm$\mmsec{0}{0005}& \decdms{-27}{25}{47}{72}$\pm$\msec{0}{01} & 329$\pm$18&${-0.0\pm0.3}$\\%300+-8 <0.24x0.1 arcsec
$[$BHB2007$]$ 19 &II & ... & ... & $<$36 &...\\
$[$BHB2007$]$ 20 &II & ... & ... & $<$36 &...\\
\bottomrule
\tabnotetext{a}{Average { integrated} flux density over the full 4.0--8.0 GHz bandwidth. Upper limits are at four times the noise level.}
  \end{tabular}
\end{table*}

\subsection{Radio sources in Lupus 1}

\begin{figure}[hb!]
\begin{center}
\includegraphics[width=0.480\textwidth,trim= 20 8 30 0, clip]{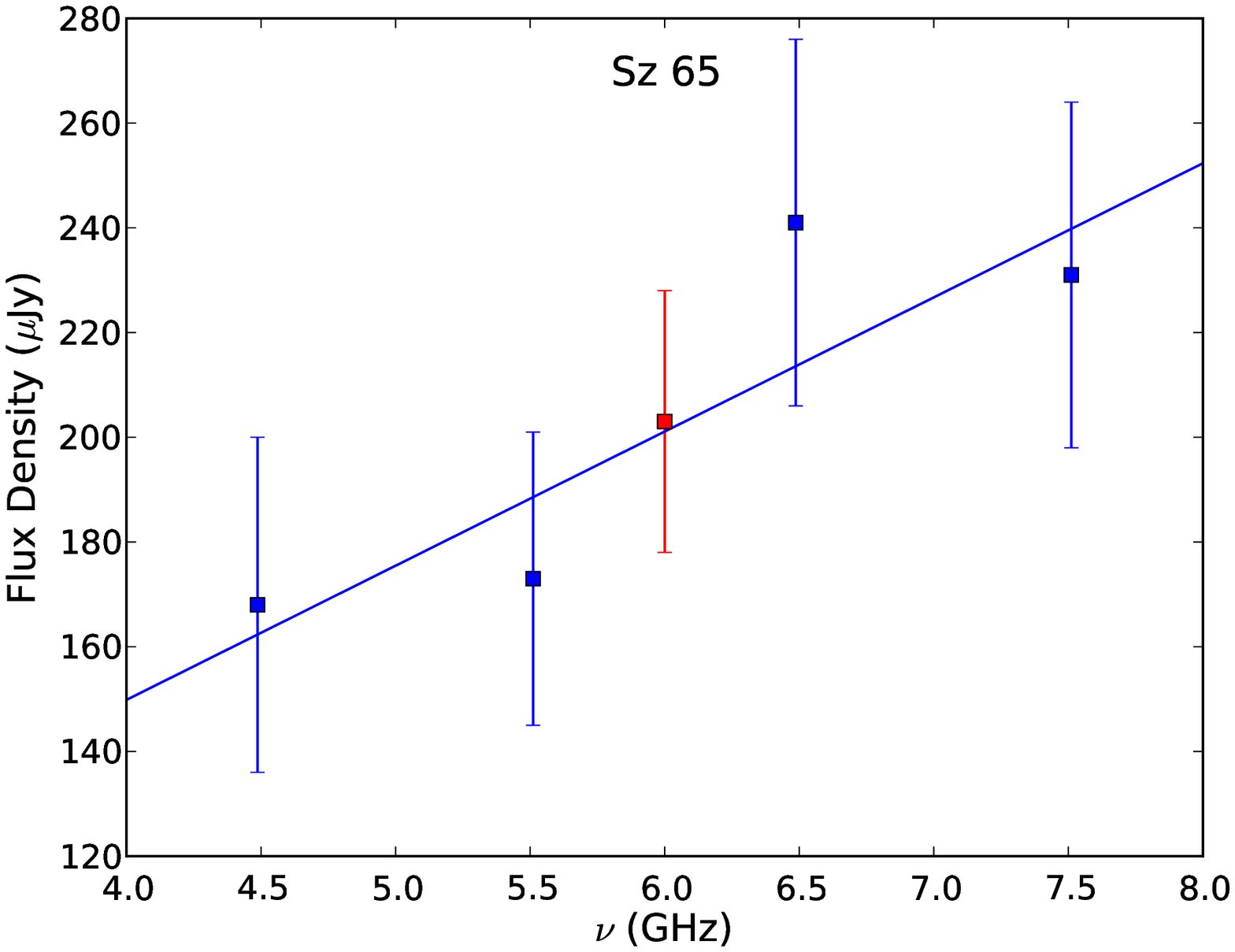}\\
\includegraphics[width=0.480\textwidth,trim= 20 8 30 0, clip]{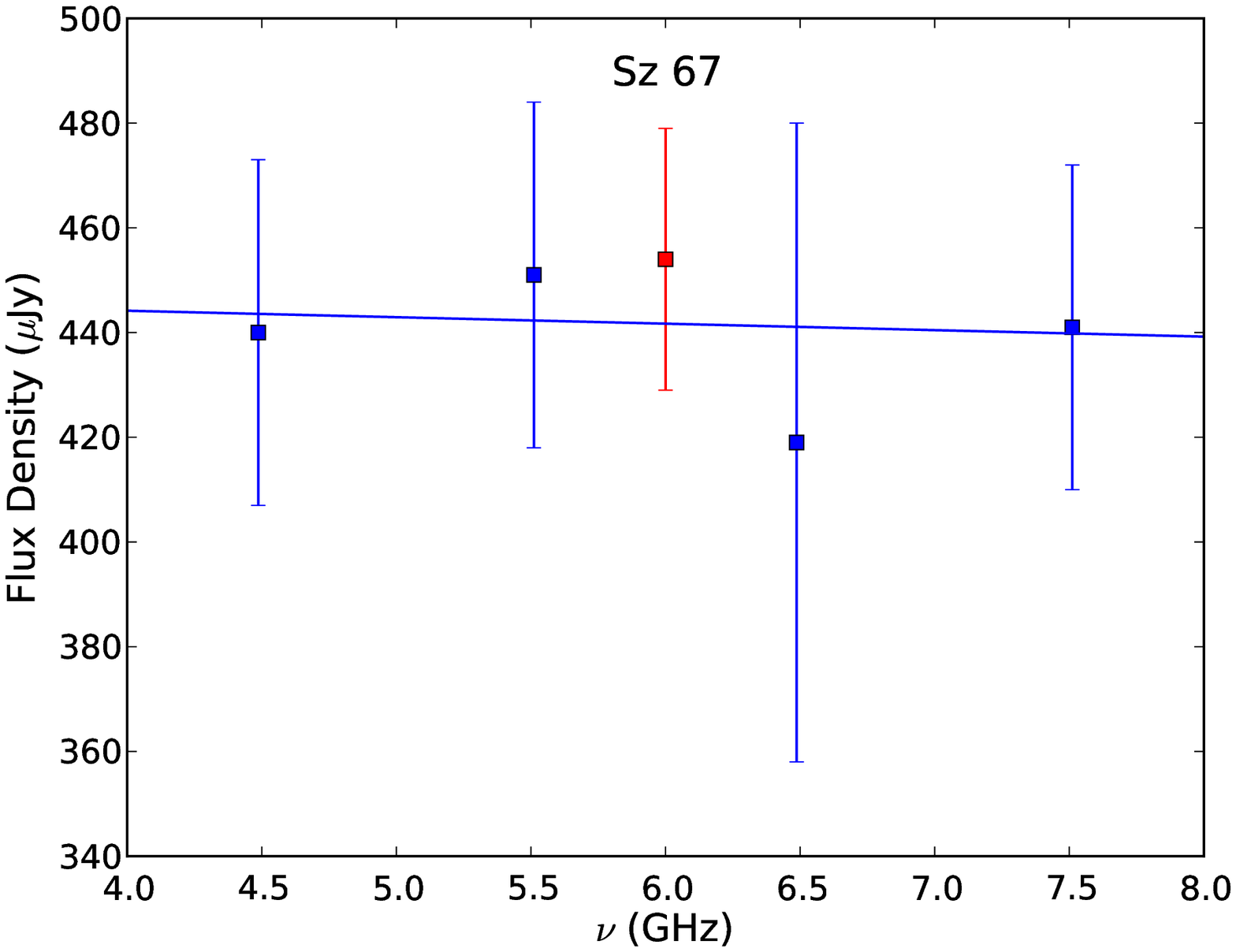}\\
\includegraphics[width=0.480\textwidth,trim= 20 8 30 0, clip]{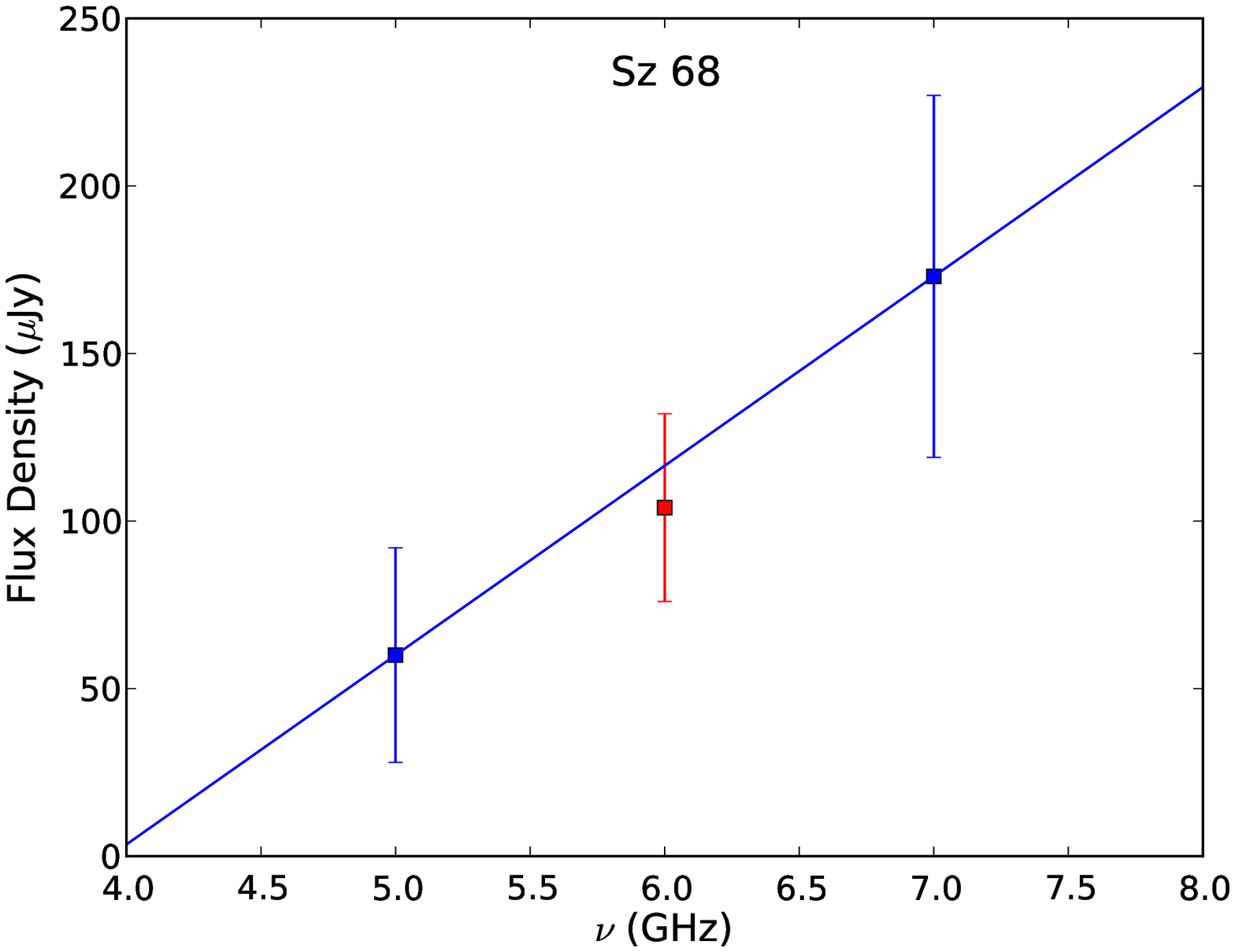}
\end{center}
\caption{Spectral index fitting for the radio sources detected in  Lupus 1 using the flux densities measured in independent sub-bands. The red point at 6~GHz is the flux density of the full data.}
\label{fig:L1_a}
\end{figure}

In the single fields corresponding to Lupus 1, a total of 58 sources were detected, four of them associated with known YSOs. This is the first time, to our knowledge that these sources are detected at radio wavelengths. The nature of the remaining sources is unclear, but most of them are likely associated to extragalactic objects, as we will discuss below. The radio properties of the detected YSOs are shown in Table~\ref{tabL1}. The deconvolved structure for all of them is consistent with point sources. Similarly to what was done in B59, the data were split in order to determine the spectral index of the detected YSOs.  The resulting spectral indices are given in Table~\ref{tabL1} and the 
fits are shown in Figure~\ref{fig:L1_a}. All the YSOs detected in Lupus 1 are close to the phase center of their observations, so the beam squint effect is expected to be negligible in this case. We searched for circular polarization (e.g., Dzib et al. 2013a) by imaging the Stokes $V$ parameter. Only one source (Sz~65) was found to show some evidence of circular polarization (see Table~\ref{tabL1}). 

\begin{table*}[!th]\centering
\footnotesize
%\small
  \setlength{\tabnotewidth}{2.0\columnwidth}
  \tablecols{9}
  \caption{Radio properties of YSOs in Lupus 1} \label{tabL1}
 \begin{tabular}{lcccccccc}
    \toprule
YSO &SED/TTS & Spect.&Ref.\tabnotemark{d} &\multicolumn{2}{c}{VLA Position} & $S_\nu\pm\sigma_{S_\nu}$\tabnotemark{a}& Circ. Pol.\tabnotemark{b} & $\alpha$\\
Name&Clas.&Clas. &  & RA    &   Dec.    &   ($\mu$Jy) & (\%) &\\
       \midrule
Sz 65 & II/CTTS&M0  & 1,2 &\dechms{15}{39}{27}{7600}$\pm$\mmsec{0}{0003} &\decdms{-34}{46}{17}{51}$\pm$\msec{0}{03} & 203$\pm$25&48$\pm$13 (L)&0.76$\pm$0.43\\%209+-12
Sz 66 &   II  &M2  & 2,3 &... & ... & $<$60 &...&...\\
Sz 67 & WTTS&M4  & 2,4 &\dechms{15}{40}{38}{2337}$\pm$\mmsec{0}{0001}& \decdms{-34}{21}{36}{89}$\pm$\msec{0}{02} & 454$\pm$25&$<$10&$-0.01\pm0.19$\\%448+-11
Sz 68 & II/CTTS&K2\tabnotemark{c} &  2,4  &\dechms{15}{45}{12}{8532}$\pm$\mmsec{0}{0006} &\decdms{-34}{17}{30}{93}$\pm$\msec{0}{07} & 104$\pm$28&$<$44&$3.1\pm1.8$\\%104+-13
Sz 69 &   II  & M1   & 2 &...  & ... & $<$64 &...&...\\
Sz 70 & CTTS    & M5   & 6,7 & ... & ... & $<$64 &...&...\\
Sz 71 & II/I    & M1.5   & 5,6  & ... & ... & $<$68 &...&...\\
Sz 72 & CTTS& M3 &1  & ... & ... & $<$60 &...&...\\
Sz 73 & CTTS& M0 &1 & ... & ... & $<$64 &...&...\\
Sz 74 & II/CTTS&M1.5&1,6 &\dechms{15}{48}{05}{2150}$\pm$\mmsec{0}{0020}& \decdms{-34}{15}{53}{00}$\pm$\msec{0}{10} & 68$\pm$32&$<$66&...\\%60+-16
Sz 75 & II/CTTS& K7\tabnotemark{c} &1,6 & ... & ... & $<$64 &...&...\\
Sz 76 & WTTS&M1  &1 & ... &... &  $<$64 &...&...\\
Sz 77 & II/CTTS&M0  &1,6 & ... & ... & $<$64 &...&...\\
\bottomrule
  \tabnotetext{a}{Average {integrated} flux density over the full 4.0--8.0 GHz bandwidth. Upper limits at four times the noise level.}
  \tabnotetext{b}{Upper limits were calculated assuming three times the noise level.}
  \tabnotetext{c}{Herbig Ae star (Herbst \& Shevchenko 1999).}
  \tabnotetext{d}{References: 1=Galli et al.\ (2015), 2=Bustamante et al.\ (2015), 3=K\"ohler et al.\ (2000), 4=Cieza et al.\ (2007), 
                 5=Alcal\'a et al.\ (2014), 6=Nuernberger et al.\ (1997) and 7= Schwartz (1977)}
  \end{tabular}
\end{table*}

\section{Discussion}

\subsection{Radio emission mechanisms in YSOs}

{The high extinction typically found along the line of sight to YSOs often prevents them from being detected at optical wavelengths. As a consequence, YSOs are usually classified on the basis of their energy distribution (SED) at infrared wavelengths.  This classification goes
sequentially from sources of Class~0 (undetected at infrared wavelengths), to Class~I, flat spectrum (FS), Class~II and Class~III  (Lada 1987; Andre et al.\ 1993 and Greene et al.\ 1994). It is intended to reflect the evolutionary sequence of the material around the YSOs. Sources of Class 0, I or FS are heavily dominated by the circumstellar envelope, sources of Class II are pre-main sequence stars surrounded by a substantial protoplanetary disk, while Class III sources are (almost) naked stars. The last two stages are usually associated with T Tauri stars, for which a different classification scheme is often used. Because they accrete material from a circumstellar disk, T Tauri stars typically exhibit some spectral lines in emission (such as H$\alpha$). T Tauri stars where this emission is more intense than a specified threshold are classified as classical T Tauri stars (CTTS), whereas those where the line emission is below that threshold are known as weak line T~Tauri stars (WTTS). Since accretion (and therefore the intensity of the emission line) is expected to decrease with the age of the young star, CTTS are thought to correspond to an earlier evolutionary stage than WTTS. Indeed, CTTS are typically associated with sources of Class II (as defined above), while WTTS are typically associated with Class III objects. It is important to emphasize, however, that the two classifications are empirical and based on different observational material.}

A significant fraction, 10$\sim$20\%, of YSOs produce radio emission by two main mechanisms. One is thermal bremsstrahlung (free-free) 
originating in partially ionized material tracing the dense base of ionized winds or jets (e.g., Rodr\'{\i}guez et al.\ 1999). The second occurs when semi-relativistic electrons gyrate in the active magnetic coronas of YSOs, producing non-thermal (gyro-synchrotron) radio emission (Feigelson \& Montmerle 1999). The different characteristics associated with these two mechanisms can be used to identify which is the dominant contribution for any particular YSO. Brightness temperature, T$_b$, is the most straightforward property that allows to distinguish between thermal (with T$_b\lesssim10^{4}$ K) and non-thermal (with T$_b\gtrsim10^{6}$ K) radio emitters. However, for sources unresolved with the VLA, only lower limits are obtained and they are often insufficient to distinguish between the two possibilities. It is known that in both mechanisms the flux density, $S_\nu$, is expected to be correlated with the frequency by a power law, characterized by the spectral index, $\alpha$. For free-free emission, it is expected that $-0.1\leq\alpha\leq+2.0$ 
(Rodr\'{\i}guez et al. 1993) with $\alpha=+0.6$ in a partially optically thick isotropic wind (Panagia 1973). For gyrosynchrotron, on the other hand, $-5.1\leq\alpha\leq+2.0$ (Dulk 1985). Thus, a very negative spectral index is indicative of non-thermal emission, but a slightly negative one is not. Moreover, flat and positive spectral indices can occur for both thermal and non-thermal emission, so other properties are required. Circularly polarized emission is a very strong indicator of non-thermal emission since free-free radiation is not expected to be polarized. Indeed, circularly polarized radio emission has been detected in some YSOs with non-thermal radio emission (e.g., G\'omez et al.\ 2008). However, many YSOs with non-thermal emission do not present measurable amounts of circularly polarized radio emission (e.g., Dzib et al.\ 2013a) because for disorganized magnetic field topologies, a practically zero net polarization could occur even for gyrosynchrotron emission.Thus an absence of polarization is not evidence for absence of non-thermal emission, and should not be taken as evidence that the radio emission is necesarily thermal. Finally, the YSOs with non-thermal radio emission are often highly variable (i.e., their flux densities can vary significantly on timescale of a few hours to a few days). In comparison, thermal YSOs show at most slow variations, on timescales of years. 

The dominant radio emission mechanism is believed to change with the evolutionary stage of the YSOs.  In the more embedded phases (Class 0, Class I, and FS), YSOs are expected to be mostly thermal emitters because the emission is dominated by the strong ionized winds driven by these very young stars. In the later (more naked) phases, the winds become weaker, and non-thermal emission associated with active coronae  tend to become dominant. Such an evolutionary trend has been suggested in the $\rho$-Ophiuchus, { Taurus, Perseus and R Corona Australis} star forming regions (Dzib et al. 2013a, 2015; Pech et al. 2015; { and Liu et al. 2014}), but it is less clear that it occurs in other regions such as Orion and Serpens (Kounkel et al. 2014; Ortiz-Le\'on et al. 2015). 

Non-thermal YSO emitters are particularly interesting because they are detectable in Very Long Base Interferometry (VLBI) observations. This enables the direct estimate of their distances, through the measurement of their trigonometric parallax. Distance is one of the most basic parameters in astrophysics, but also, unfortunately, one of the most difficult to measure, especially for YSOs that are hard to detect at optical wavelengths. VLBI observations of non-thermal YSOs have been instrumental to improve the situation (i.e.,  Loinard et al.\ 2005, 2008; Menten et al.\ 2007; Dzib et al.\ 2010, 2011). A major project, the Gould's Belt Distances Survey, has been initiated with the objective of measuring the trigonometric parallax and proper motions of roughly 200 YSOs distributed in the five most often studied nearby star-forming regions ($\rho$-Ophiuchus, Taurus, Perseus, Orion and Serpens; see Loinard et al. 2011 for a discussion). A study of the radio emission of YSOs in other nearby star forming regions, and subsequent VLBI observations, could help expand the objectives of the Gould's Belt Distances Survey. Indeed, the distances to B59 and Lupus 1 remain very uncertain, as we now discuss. 

The estimated distances for the Pipe Nebula are 130$^{+24}_{-58}$ pc (using parallax determinations from field stars in the Hipparcos catalog; Lombardi et al.\ 2006) and 145 $\pm$ 16 pc (through polarimetry of field stars from the Hipparcos catalog; Alves \& Franco 2007). However, there is no distance estimation to B59 itself, and it is usually assumed to be at 130 pc, the most common adopted distance to the Pipe Nebula. The range of distances estimated for Lupus 1, on the other hand, vary from 100 pc to 300 pc. Comer\'on (2008) argued that a single value of the distance is probably inadequate. Particularly, the Lupus 3 cloud seems to be at about 200 pc, while for the rest of the clouds a value of 150 pc is more adequate. However, this seems to disagree with the most recent results by Galli et al.\ (2013) who studied kinematic parallaxes of T Tauri stars in Lupus and found that the distances for Lupus 1 and Lupus 3 are 182$^{+7}_{-6}$ and 185$^{+11}_{-10}$, respectively. Detecting non-thermal candidates in these regions and subsequently measuring their trigonometric parallax through VLBI observations could significantly reduce the uncertainties on the distance to both regions.

\subsection{Radio emission of YSOs in B59 and Lupus 1}

The spectral indices reported here for detected YSOs in B59 and Lupus 1 do not allow to clearly distinguish between a thermal and a non-thermal origin. The source $[$BHB2007$]$~7 has a negative spectral index, but given the large errors, this does not provide an unambiguous
identification for the emission mechanism. The circular polarized emission detected from Sz~65, on the other hand, clearly demonstrate that the radio emission in this source is non-thermal. The presence of circularly polarized emission is detected at the 4$\sigma$ level in the $V$ Stokes image (F$_c$ = 98 $\pm$ 24 $\mu$Jy), resulting in the 48 $\pm$ 13 \% level of circular polarization mentioned in Table 3. Other properties have to be taken into account in order to favor a mechanism for the observed radio emission of the rest of the YSOs.

To roughly study variability in B59, a comparison between the flux densities presented in Table~\ref{tabB59} and those reported by Dzib et al. (2013b), plotted as green squares in Figure~\ref{fig:b59_a}, can be performed. The radio fluxes are consistent whithin 1$\sigma$ for the sources $[$BHB2007$]$~10 and $[$BHB2007$]$~11, but they are not consistent for the other three YSOs detected by Dzib et al. (2013b), which all have lower fluxes in the observations reported here than in those described by Dzib et al.\ (2013b). This apparently systematic effect is suspicious since the variations due to magnetic activity ought to be random. We note that the new observations have a higher angular resolution, so it is not clear if the variation is due to a real variation or to part of the extended radio emission being resolved out.

An alternative approach to distinguish between thermal and non-thermal radio emission is to use the relation that the radio emission is more thermal in origin for younger YSOs and more non-thermal for evolved YSOs {(e.g., Dzib et al. 2013a, 2015; Pech et al. 2015;  and Liu et al. 2014)}. Given the early SED classification of the sources in B59, these are expected to be mostly thermal radio emitters, as also discussed by Dzib et al.\ (2013b). The situation for Lupus 1 is similar, as most of  the detected YSOs are of Class II and CTTS. The only exception is the WTTS Sz~67, which might  more likely be a non-thermal emitter. However, this would clearly have to be examined in more detail with further observations aimed at constraining its variability and/or 
its brightness temperature.

Finally, the detection rate of YSOs at radio frequencies is 10--20\% (e.g., Andr\'e 1996). From our observations, we conclude that in B59 and Lupus 1, the detection rates are 45\% and 30\%, respectively. The main factor that helps to obtain higher detection rates is the higher sensitivities of the present observations as compared to similar surveys realized prior to the era of the upgraded  VLA (e.g., O'Neal et al.\ 1990, Gagne et al.\ 2004). Also, the distance to the star forming region can play a significant role, since the radio flux of stars will be systematically dimmer at farther distances. In B59, for instance, the high detection rate is probably a combination of the increased sensitivity and the proximity of the region.

\subsection{Background Objects}

\begin{table*}[!ht]\centering
\small
  \tablecols{6}
  
  % Stretch the space between table columns 
  %\setlength{\tabcolsep}{2.8\tabcolsep}
  \caption{Other radio sources detected in the B59 region} \label{tab:b59_bg}
 \begin{tabular}{cc|cc|cc}
    \toprule
VLA &$S_\nu\pm\sigma_{S_\nu}$&VLA &$S_\nu\pm\sigma_{S_\nu}$&VLA &$S_\nu\pm\sigma_{S_\nu}$\\
Name&       (mJy)        &Name&        (mJy)       &Name&(mJy)\\
       \midrule
J171052.52-272322.1&0.568$\pm$0.039&J171130.18-272921.6&0.507$\pm$0.028&J171149.76-272414.0&0.075$\pm$0.012\\
J171053.90-273141.6&0.124$\pm$0.029&J171132.66-271959.0&0.249$\pm$0.018&J171150.08-273219.0&0.071$\pm$0.017\\
J171054.36-273129.1&0.114$\pm$0.031&J171133.70-272757.1&0.264$\pm$0.017&J171152.18-273418.7&0.448$\pm$0.038\\
J171054.58-273047.5&0.103$\pm$0.028&J171134.14-271720.8&0.666$\pm$0.042&J171152.96-271706.4&0.160$\pm$0.031\\
J171056.08-273231.1&0.117$\pm$0.032&J171137.05-273356.5&0.131$\pm$0.024&J171153.64-271651.1&0.433$\pm$0.040\\
J171057.78-271957.7&0.151$\pm$0.025&J171138.16-273242.4&0.069$\pm$0.017&J171154.72-272209.9&1.616$\pm$0.082\\
J171058.70-272605.2&0.359$\pm$0.026&J171138.36-273205.7&0.113$\pm$0.016&J171155.35-272102.7&0.068$\pm$0.014\\
J171058.77-272854.1&0.074$\pm$0.018&J171139.56-273315.6&0.070$\pm$0.020&J171157.72-271956.5&0.063$\pm$0.018\\
J171059.63-271949.0&0.084$\pm$0.024&J171140.53-273237.1&0.057$\pm$0.016&J171159.10-272439.7&0.655$\pm$0.036\\
J171101.02-272450.3&0.379$\pm$0.025&J171140.81-272310.2&0.115$\pm$0.012&J171159.27-271735.5&0.215$\pm$0.032\\
J171101.61-272546.1&0.057$\pm$0.016&J171141.14-273327.9&0.722$\pm$0.041&J171202.52-273217.5&0.325$\pm$0.027\\
J171102.21-272115.9&0.072$\pm$0.016&J171143.63-273400.9&2.050$\pm$0.110&J171202.97-272325.4&0.066$\pm$0.017\\
J171104.20-272341.8&0.054$\pm$0.014&J171143.74-271823.7&0.082$\pm$0.020&J171205.87-272737.8&8.527$\pm$0.427\\
J171105.05-272117.4&0.053$\pm$0.015&J171143.86-271825.0&3.203$\pm$0.161&J171206.52-272501.5&0.098$\pm$0.019\\
J171107.05-273145.1&0.108$\pm$0.018&J171144.17-271833.3&0.278$\pm$0.024&J171209.68-271922.7&0.135$\pm$0.036\\
J171113.35-273327.8&0.110$\pm$0.023&J171144.59-273335.4&7.884$\pm$0.395&J171210.23-273120.4&0.172$\pm$0.031\\
J171113.42-272738.8&0.120$\pm$0.013&J171145.44-271857.2&0.163$\pm$0.019&J171210.79-272108.4&0.183$\pm$0.032\\
J171119.52-272101.4&0.051$\pm$0.013&J171145.45-271856.2&0.144$\pm$0.018&J171212.13-273012.3&0.164$\pm$0.032\\
J171123.81-272815.6&0.061$\pm$0.011&J171145.81-271711.2&0.473$\pm$0.037&J171212.46-272615.9&0.200$\pm$0.031\\
J171126.62-272615.9&0.065$\pm$0.011&J171146.30-273305.7&0.114$\pm$0.024&J171212.38-272225.2&0.179$\pm$0.032\\
J171127.29-272404.5&0.077$\pm$0.012&J171147.00-273149.7&1.022$\pm$0.053&\\
\bottomrule
  \end{tabular}
\end{table*}

\begin{table*}[!ht]\centering
\small
  \tablecols{6}
  
  % Stretch the space between table columns 
  %\setlength{\tabcolsep}{2.8\tabcolsep}
  \caption{Other radio sources detected in the Lupus 1 region} \label{tab:l1_bg}
 \begin{tabular}{cc|cc|cc}
    \toprule
VLA &$S_\nu\pm\sigma_{S_\nu}$&VLA &$S_\nu\pm\sigma_{S_\nu}$&VLA &$S_\nu\pm\sigma_{S_\nu}$\\
Name&       (mJy)        &Name&        (mJy)       &Name&(mJy)\\
       \midrule
J153913.43-344642.1&0.318$\pm$0.029&J154516.47-342246.9&1.085$\pm$0.077&J154921.75-354620.0&0.329$\pm$0.039\\
J153925.70-344903.0&0.134$\pm$0.022&J154522.87-341726.5&0.204$\pm$0.020&J154922.64-354010.3&0.143$\pm$0.021\\
J153927.13-344923.9&0.149$\pm$0.027&J154640.39-342834.2&0.191$\pm$0.020&J154931.02-355218.5&0.274$\pm$0.025\\
J153933.01-344530.3&0.297$\pm$0.022&J154641.06-343111.0&0.138$\pm$0.018&J154932.47-353600.2&0.343$\pm$0.063\\
J153935.71-344307.0&1.200$\pm$0.066&J154642.79-342642.7&0.170$\pm$0.032&J154939.04-354558.4&0.384$\pm$0.045\\
J153943.99-344223.3&1.156$\pm$0.086&J154656.69-343423.3&0.285$\pm$0.054&J154943.05-354529.2&0.405$\pm$0.064\\
J153944.64-344628.0&0.558$\pm$0.040&J154700.04-343453.1&2.664$\pm$0.161&J154950.21-355419.9&0.635$\pm$0.119\\
J154027.16-342301.8&0.819$\pm$0.047&J154738.47-351705.7&0.297$\pm$0.055&J155135.82-360110.5&6.390$\pm$0.326\\
J154028.33-341938.4&15.319$\pm$0.766&J154739.08-351658.6&0.290$\pm$0.053&J155135.86-360106.9&32.649$\pm$1.634\\
J154028.48-341938.4&0.182$\pm$0.026&J154742.83-353000.7&0.674$\pm$0.039&J155136.19-360122.7&1.846$\pm$0.114\\
J154028.69-341938.2&5.882$\pm$0.295&J154750.81-352605.3&0.460$\pm$0.031&J155136.60-355204.7&0.918$\pm$0.078\\
J154043.48-341855.9&0.484$\pm$0.033&J154800.51-351544.3&0.091$\pm$0.017&J155141.74-355310.9&4.936$\pm$0.249\\
J154048.51-341954.0&0.121$\pm$0.023&J154811.61-352437.6&0.534$\pm$0.105&J155145.34-360123.2&1.017$\pm$0.070\\
J154054.98-342529.1&0.764$\pm$0.075&J154852.60-353827.4&0.176$\pm$0.036&J155149.83-360125.6&0.783$\pm$0.064\\
J154102.29-341843.7&0.594$\pm$0.101&J154903.88-353838.7&0.125$\pm$0.019&J155149.88-355712.8&0.825$\pm$0.044\\
J154453.70-341358.8&0.910$\pm$0.129&J154914.18-355341.1&0.344$\pm$0.067&J155149.89-355724.5&0.091$\pm$0.017\\
J154500.26-341856.4&0.125$\pm$0.026&J154916.62-354015.9&0.165$\pm$0.020&J155150.22-360118.5&0.292$\pm$0.050\\
J154506.19-342131.0&0.175$\pm$0.036&J154917.36-353545.8&0.138$\pm$0.028&J155202.39-355932.0&0.417$\pm$0.043\\
J154511.41-341700.2&0.089$\pm$0.017&J154917.87-353938.9&0.152$\pm$0.019&\\
J154515.92-342224.9&0.403$\pm$0.047&J154920.46-353448.5&0.247$\pm$0.047&\\
\bottomrule
  \end{tabular}
\end{table*}

Several sources were detected in the B59 (Table~\ref{tab:b59_bg}) and Lupus 1 (Table~\ref{tab:l1_bg}) regions without any clear counterpart at other wavelengths. To reflect the fact that these  sources were found using the VLA, a source with coordinates {\it hhmmss.ss$+$ddmmss.s} will be named as VLA J{\it hhmmss.ss$+$ddmmss.s}. A brief discussion on their nature follows. 

Fomalont et al.\,(1991), by observing an area free of Galactic objects, found that the number of expected background extragalactic sources per arcmin$^2$ with a flux larger than $S$ and observing at 5~GHz is given by:

\begin{displaymath}
 \left(\frac{N}{arcmin^2}\right)=0.42\pm0.05\left(\frac{S}{30 \mu Jy}\right)^{-1.18\pm0.19}.
\end{displaymath}

The observed area in the B59 mosaic is { 316 arcmin$^2$.} Assuming that the noise is uniform in the entire map, above 45 $\mu$Jy (5$\times\sigma_{\rm noise}$) { a total of 82$\pm$16 } background sources are expected.  This estimate strongly suggests that the 62 sources in Table~\ref{tab:b59_bg} are extragalactic objects. {The slightly smaller number of detected sources in B59 might in part be due to the non uniformity of the noise pattern in the image, which are noisier near their edges. It could also simply reflect slight cosmic variance.} On the other hand, for the individual fields in Lupus 1, the noise follows a gaussian attenuation pattern corresponding to the primary beam response of the VLA antennas. Taking this into account, Equation A11 from Anglada et al. (1998) is adopted to obtain that the number $N$ of expected background sources above a flux density $S$, at the observed frequency, in each field:

\begin{displaymath}
 N=0.85\left(\frac{S}{\rm mJy}\right)^{-0.75}.
\end{displaymath}

\noindent
This yields a number of expected background sources in Lupus 1 of 54$\pm$7. This number is in good agreement with the 58 sources detected without counterpart at other wavelengths (Table~\ref{tab:l1_bg}), suggesting that they are all of extragalactic origin.

\section{Conclusions}

Deep C-band radio observations ($\sigma$ $<15\,\mu$Jy) of the B59 and Lupus 1 regions were presented. In the B59 region, nine radio sources associated with YSOs were detected. Five were previously known to be radio emitters and the remaining four are reported here for the first time. On the other hand, four radio sources associated with YSO were detected (here for the first time) in the Lupus 1 region. The properties of the radio emission from most of the detected YSOs are consistent with a thermal origin. The $[$BHB2007$]$~7 star is the only source whose (marginally detected) circular polarized emission suggests a non-thermal origin. Additionally, a non-thermal radio emission is suggested as plausible for the case of the WTTS Sz~67 on the basis of the evolutionary status (Class III/WTTS) of the star. However, further observations are clearly required to test this hypothesis, by measuring the variability and/or brightness temperature of this radio source.
In addition to the YSO mentioned here, several tens of radio sources are detected in both regions. They are all most likely extragalactic background sources seen through the B59 and Lupus 1 star-forming regions.

\acknowledgments
L.L. and L.F.R. acknowledge the financial support of DGAPA, UNAM, and CONACyT, M\'exico. This research has made use of the SIMBAD database, operated at CDS, Strasbourg, France

\end{document}